\documentclass[12pt]{book}
\usepackage[T1]{fontenc}

\usepackage{betadecay}

\begin{document}

\title{X-RAY BURSTS AND PROTON CAPTURES CLOSE TO THE DRIPLINE}

\author{\underbar{T.~Rauscher}\Iref{1}, F.~Rembges\Iref{1}, H.~Schatz\Iref{2},
M.~Wiescher\Iref{3}, F.-K.~Thielemann\Iref{1} }

\maketitle

\Instfoot{1}{Department of Physics and Astronomy, University of Basel,
4056 Basel, Switzerland} \Instfoot{2}{GSI, Darmstadt, Germany}
\Instfoot{3}{Department
of Physics, University of Notre Dame, Notre Dame, IN, USA}

\index{Rauscher~T.}
\index{Rembges~F.}
\index{Schatz~H.}
\index{Wiescher~M.}
\index{Thielemann~F.-K.}

\begin{abstract}
The hydrogen-rich accreted envelopes of neutron stars in binary systems
are the site for the rapid proton capture nucleosynthethic process (\emph{rp}
process), which involves nuclei close to the proton dripline. An overview
of the relevant reactions and nuclear properties for \emph{rp}-process
studies is given, along with motivations of further experimental nuclear
physics studies.
\end{abstract}

\section{Introduction}

In low-mass binary systems involving a neutron star, proton-rich material
from the atmosphere of the companion giant star can be accreted on the
surface of the neutron star
\cite{rauscher:woo76,rauscher:mar77,rauscher:jos77,rauscher:wall81}.
(see also the review article \cite{rauscher:lew93}). Once a critical density
(\( \approx 10^{6} \) g/cm\( ^{3} \)) is reached within the accreted
layer, thermonuclear fusion of hydrogen and helium is ignited and high
temperatures are attained in an explosive runaway. Such thermonuclear flashes
can be observed as so-called type I X-ray bursts. The energy is generated
by hot hydrogen burning cycles and by burning helium in the 3\( \alpha  \)
reaction and a sequence of (\( \alpha  \),p) and (p,\( \gamma  \)) reactions,
which provide seed nuclei for the subsequent hydrogen burning in the \emph{rp}
process, consisting of rapid proton captures and \( \beta  \)\( ^{-} \)
decays close to the proton dripline. The timescale of the \emph{rp} process
is given by the slow electron capture half-lives (waiting points) in the
process path.

For a long time the doubly magic nucleus \( ^{56} \)Ni (\( t_{1/2}=2\times
10^{4} \)
s) was considered to be the endpoint of the \emph{rp} process. The reaction
network was extended up to Sn in a recent one-zone model calculation
\cite{rauscher:schatz98},
and it was found that \( ^{56} \)Ni only becomes a temporary waiting point
at the initial rise of the burst but that the reaction flow can go beyond
in the cooling phase. Thus, nuclei in the mass range \( A\approx 80-100 \)
can be synthesized within the short timescale (\( \approx 10-100 \) s)
of the explosive event. This results in an enhanced energy production,
additional hydrogen consumption and altered ashes of the \emph{rp} process,
which are deposited on the crust of the neutron star. If mass ejection
during the burst is feasible, this would also open up a new possible
explanation
of the so-called \emph{p} nuclei in the solar abundance pattern, which
are stable proton-rich nuclei which cannot be produced in the \emph{s}
and \emph{r} processes.

Of particular importance proved to be two-proton capture reactions which
can bridge the waiting points encountered at the \( N=Z \) nuclei \( ^{64}
\)Ge(\( t_{1/2}=64 \)
s), \( ^{68} \)Se(\( t_{1/2}=36 \) s), and \( ^{72} \)Kr(\( t_{1/2}=17 \)
s).

\begin{center}
\includegraphics*[height=10cm]{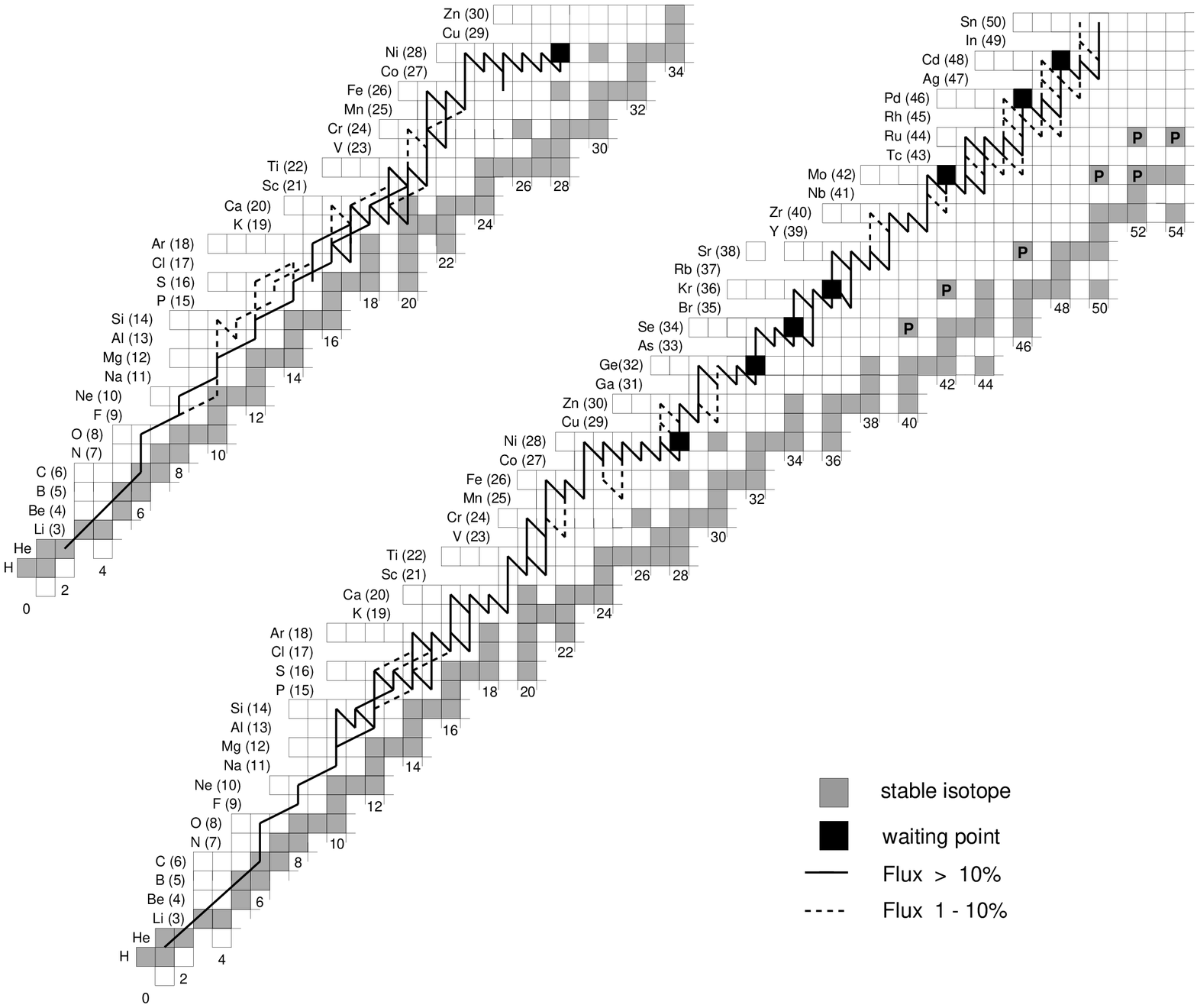}
\betacaptionfig{The \emph{rp}-process reaction-flux integrated
over the thermonuclear runaway (top) and the cooling phase (bottom) of
a single X-ray burst \cite{rauscher:schatz98}. Stable \emph{p} nuclei
(which cannot be produced in the s and r process) are marked by \textbf{P}.}
\betalabelfig{rauscher:rp-path}
\end{center}

Time-dependent calculations with coupled full networks are currently performed
to give a more consistent understanding of the time structure of burst
and interburst phases, and of the fuel consumption
\cite{rauscher:remoak,rauscher:rem98,rauscher:remdiss}.

\section{Proton capture reactions}

The reaction path of the \emph{rp} process is shown in
Fig.~\ref{rauscher:rp-path}.
Similar to the \emph{r} process, far away from stability the \emph{rp}-process
path is determined by the nuclear masses or proton separation energies,
respectively, and the \( \beta  \)\( ^{-} \)-decay half-lives alone and
not by individual reaction rates. Only several individual (p,\( \gamma  \))
and (p,\( \alpha  \)) reactions in the early burst phase and the 2p-capture
reactions \( ^{56} \)Ni(2p,\( \gamma  \))\( ^{58} \)Zn, \( ^{64} \)Ge(2p,\(
\gamma  \))\( ^{66} \)Se,
\( ^{68} \)Se(2p,\( \gamma  \))\( ^{70} \)Kr, and \( ^{72} \)Kr(2p,\( \gamma 
\))\( ^{74} \)Y
will have major impact on the resulting reaction flow, energy generation,
and nucleosynthesis. Although the 2p-capture rates are slow due to the
short lifetime of the proton-unbound intermediate nucleus after the first
capture, they are essential for the continuation of the reaction path because
they compete with quite slow \( \beta  \)\( ^{-} \) decays. Current estimates
of these rates at the waiting points are based on mass models and level
density calculations. As the rates and derived lifetimes against proton
capture depend sensitively on the reaction \emph{Q} value, more experimental
data about nuclear masses in this region are clearly needed to confirm
the results. 
For instance, the stellar lifetime (including \( \beta  \)
decay and proton capture) of \( ^{68} \)Se is 
\begin{center}
\includegraphics*[height=6.0cm]{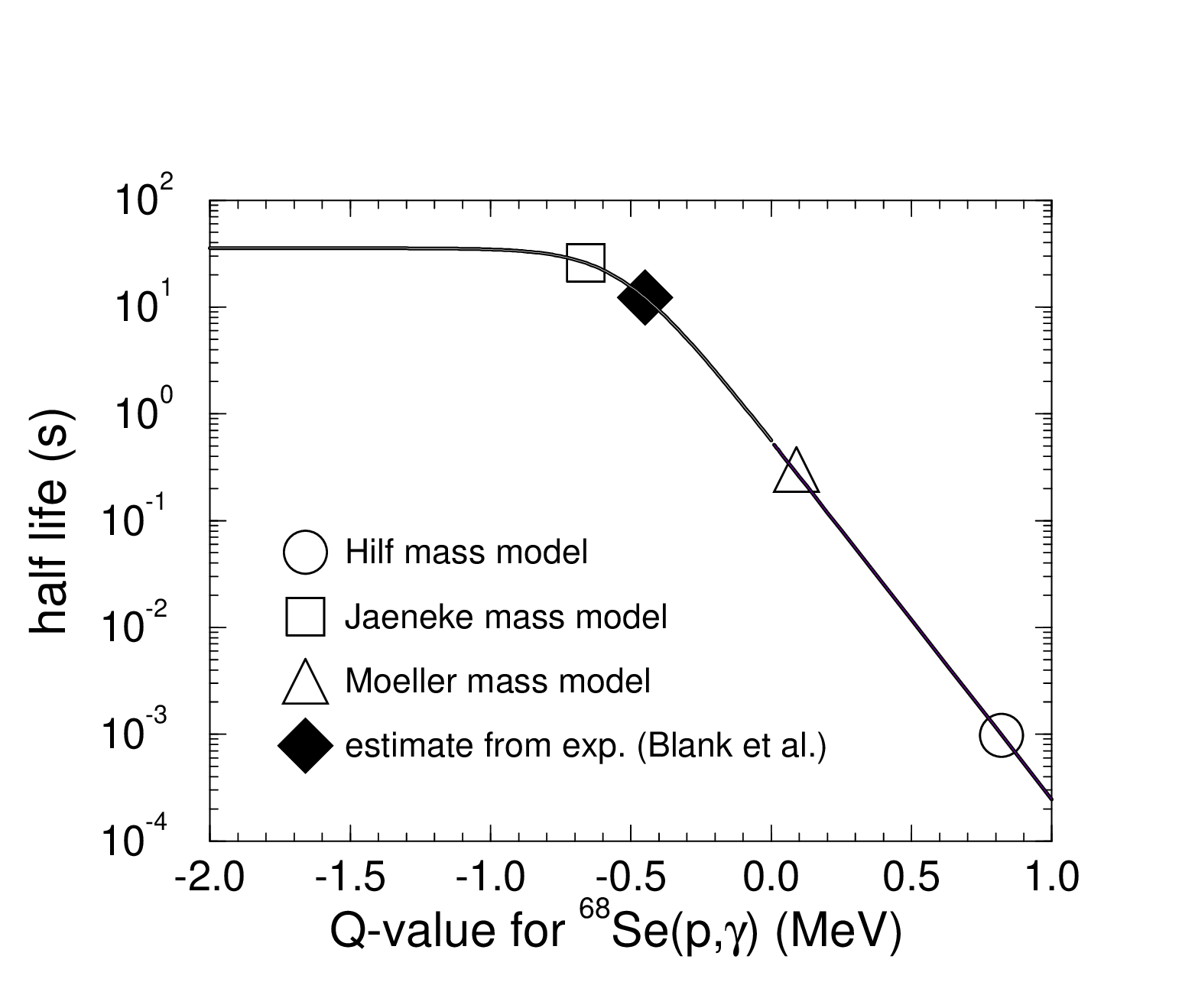}
\betacaptionfig{Stellar halflife (including \protect\( \beta \protect \)
decay \emph{and} p-capture) of \protect\( ^{68}\protect \)Se as a function
of reaction \emph{Q} value for a temperature of 1.5 GK, a density of
10\protect\( ^{6}\protect \)
g/cm\protect\( ^{3}\protect \), and solar hydrogen abundance
\cite{rauscher:schatz98}.
\emph{Q} values predicted by different mass models are indicated.}
\betalabelfig{rauscher:se68}
\end{center}
shown in Fig. \ref{rauscher:se68}
as a function of the proton capture \emph{Q} value. A change in the \emph{Q}
value of only 200 keV, well within mass model uncertainties, might change
the stellar lifetime by a factor of 5. Put in another way: For a lifetime
determination to better than a factor of 2, the \emph{Q} value has to be
known with an accuracy of better than 100 keV, which is way beyond the
accuracy of modern mass model predictions.
While a direct study of 2p-capture reactions is not possible, the study
of Coulomb dissociation of \( ^{66} \)Se, \( ^{70} \)Kr, and \( ^{74} \)Y
offers a way to obtain information about the reactions \cite{rauscher:kaepp98}.

In the early burst phase, hot hydrogen cycles are formed which consist
of two subsequent proton captures, a \( \beta  \) decay, another proton
capture and \( \beta  \) decay, and a final (p,\( \alpha  \)) reaction
closing the cycle. With rising temperature, these cycles break up by (p,\(
\gamma  \))
reactions in the order of decreasing \emph{Q} values. Most of these rates
can be predicted in the statistical model of nuclear reactions, as the
prerequisite of a sufficient number of resonances within the Gamow window
is met \cite{rauscher:rtk97,rauscher:rem98}. Only at the low proton separation
energies close to the dripline, the level density becomes too low to justify
averaging over resonances and the contribution of isolated resonances has
to be taken into account. Nevertheless, measurements of reactions are highly
desireable in all cases, in order to obtain information on the level density
and to verify the theoretical calculations
\cite{rauscher:schatz98,rauscher:adndt99}.

Recently, a study on the influence of proton capture
on \( ^{27} \)Si, \( ^{31} \)S, \( ^{35} \)Ar, and \( ^{39} \)Ca
on hot hydrogen burning became available \cite{rauscher:ili99}.
Estimates of the rates are given, based on previously published values
or updated nuclear properties.
The rate used for \( ^{31} \)S(p,\( \gamma  \))\( ^{32} \)Cl 
\cite{rauscher:vouz94} is more than
a factor of 3 slower than the one previously used \cite{rauscher:wor94}
in \emph{rp}-process calculations, whereas the other rates differ by about
30\%. (Note, however, that the \( ^{31} \)S(p,\( \gamma  \)) rate is also in
disagreement with another recent evaluation \cite{rauscher:lef97}.)
A self-consistent time-dependent 
calculation 
\cite{rauscher:remdiss,rauscher:remprep}
shows the importance of these rates in the initial phase of the burst
\begin{center}
\includegraphics*[height=8.0cm]{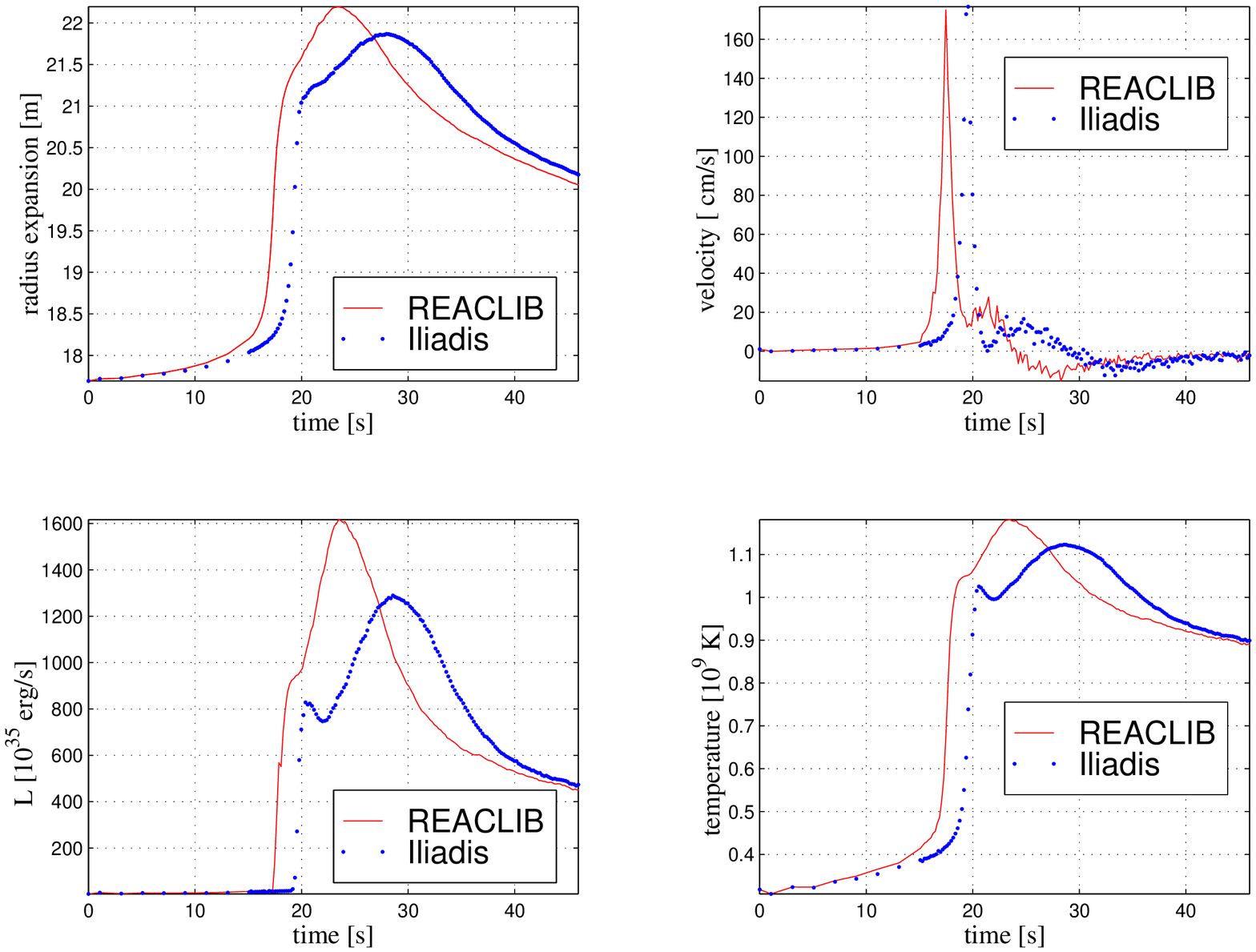}
\betacaptionfig{Comparison of burst profiles with previous rates
(REACLIB) \cite{rauscher:wor94} and the rates given in
\cite{rauscher:ili99} (Iliadis) in a self-consistent model
\cite{rauscher:remdiss,rauscher:remprep}.
Due to the slower rates, the onset of the burst is delayed, reflected
in a delay in the expansion, the luminosity and the temperature profile.
Energy generation is suppressed, shown by the lower maxima
in luminosity and temperature.}
\betalabelfig{rauscher:ilicomp}
\end{center}
(Fig.~\ref{rauscher:ilicomp}).
Due to the slower rates, the onset of the thermonuclear runaway is delayed
and the luminosity and temperature profiles are altered.
The strongest effect is due to the \( ^{31} \)S(p,\( \gamma  \))\( ^{32} \)Cl
rate
which is also reflected in the development of the abundances over time,
as shown in Fig.~\ref{rauscher:iliabu}. This is in contradiction with
the (not self-consistent) post-processing study claiming to have found no
significant impact \cite{rauscher:ili99}.
Together with the nuclear physics
uncertainties involved, it proves the importance
of measuring such reactions in radioactive ion beam facilities.

\section{\protect\( \beta \protect \) decays}

As mentioned before, the rp-process flow is determined only by the \emph{Q}
values and \( \beta ^{-} \)-decay half-lives as long as the temperature
is high enough to permit a (p,\( \gamma  \))--(\( \gamma  \),p) equilibrium.
Then the processing to heavy nuclei is determined by the lifetime of the
waiting point nuclei. The \( \beta  \)-decay half-lives of the critical
waiting point nuclei are well-known, even when the proton capture \emph{Q}
values are not. The few unknown half-lives encountered below Zr are those
of the nuclei reached by 2p captures. Their half-lives will have little
impact since the reaction flow is determined by the slow 2p captures. Only
beyond Zr, the \emph{rp}-process path fully enters a region of unmeasured
\( \beta  \)-decay half-lives. Usually, theoretical half-lives from a
certain model are adopted in \emph{rp}-process studies, e.g. from QRPA
\cite{rauscher:qrpa}, supplemented by shell model 
calculations \cite{rauscher:oxbash}
in the one-zone model \cite{rauscher:schatz98} quoted above. Masses also
enter crucially. When comparing predictions of different models it is important
to not only consider the average 
\begin{center}
\includegraphics*[height=8.0cm]{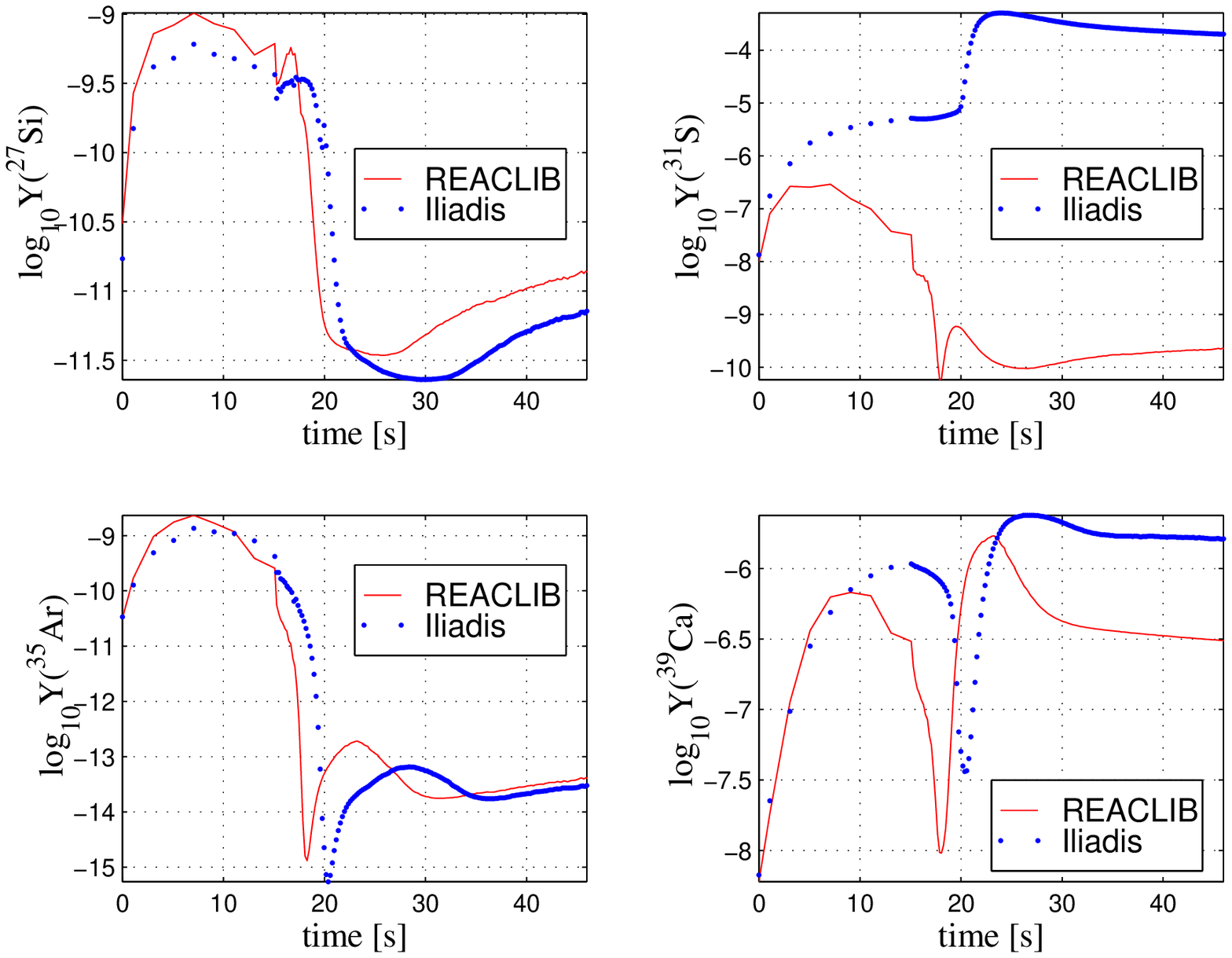}
\betacaptionfig{Abundances of \protect\( ^{27}\protect \)Si,
\protect\( ^{31}\protect \)S, \protect\( ^{35}\protect \)Ar, and \protect\(
^{39}\protect \)Ca
vs. time. Compared are self-consistent calculations
\cite{rauscher:remdiss,rauscher:remprep}
with previous rates (REACLIB) \cite{rauscher:wor94} and with rates from
\cite{rauscher:ili99} (Iliadis). The strongest effect is due to
the \protect\( ^{31}\protect \)S(p,\protect\( \gamma \protect \))\protect\(
^{32}\protect \)Cl
rate.}
\betalabelfig{rauscher:iliabu}
\end{center}
global reliability but also the behavior
of the uncertainties when the model is extrapolated towards proton-rich
nuclei \cite{rauscher:schatz98}.

\( \beta  \)-decay half-lives in high temperature scenarios can be altered
in respect to laboratory values due to the decay of thermally excited states.
It has been shown \cite{rauscher:qrpa} that the \( \beta  \)-decay lifetime
of excited states can be significantly different from the lifetime of the
ground state. In the \emph{rp} process, this effect could mainly be important
for the long lived self-conjugate nuclei, forming the waiting points in
the process path. For these nuclei only the population of the first excited
2\( ^{+} \) state plays a role in the relatively low energy regime of
\( kT\leq 300 \) keV of X-ray bursts. The energies of the 2\( ^{+} \)
states were estimated in a valence scheme and with shell model calculations
\cite{rauscher:schatz98}. Significant deviations from the ground state
decay are only found for \( ^{64} \)Ge, \( ^{68} \)Se, and \( ^{72} \)Kr.
However, for temperatures below 2 GK the decay rates stay constant with
temperature and around 3 GK the effect is still less than approximately
15\% and thus negligible. Only in scenarios with temperatures exceeding
3 GK temperature dependence of \( \beta  \)-decay half-lives has to be
taken into account.

\section{Conclusions}

Rapid proton captures along the proton dripline are typical for nucleosynthesis
processes at the high density and temperature conditions of an accreted
neutron star envelope. Recent investigations have shown that the \emph{rp}
process proceeds well beyond Ni and can produce nuclei in the mass range
\( A\approx 80-100 \). To establish the definite endpoint of the \emph{rp}
process, further experimental investigations of nuclear properties (masses,
\( \beta  \)\( ^{-} \)-decay half-lives, reaction rates) at or close
to the dripline are needed. In the lower mass range, reactions of the hot
hydrogen burning cycles determine the initial burst phase and the timescale
of the processing into more heavy nuclei. Thus, they are interesting targets
for current and future studies at radioactive ion beam facilities. For
an extensive review on experimental needs and approaches in the \emph{rp}
process and in astrophysics in general, see \cite{rauscher:kaepp98}. Consistent
multizone, hydrodynamical calculations are important to consider mixing
between different layers and to study the influence of the reaction rates
on fuel consumption, energy generation and nucleosynthesis in X-ray bursts.

\section*{Acknowledgements}

This work was supported by the Swiss National Science Foundation (grant
2000-053798.98).

\end{document}